\documentclass[aps,prb,twocolumn,groupedaddress,amsmath,amssymb]{revtex4}

\usepackage{graphicx}
\usepackage{hyperref}
\usepackage{pifont}
\usepackage{color}

\def\mean#1{\left< #1 \right>}

\begin{document}

\title{Quantum-confinement and Structural Anisotropy result in Electrically-Tunable Dirac Cone in Few-layer Black Phosphorous}

\author{Kapildeb Dolui and Su Ying Quek}
\affiliation{Department of Physics, Centre for Advanced 2D Materials and Graphene Research Centre, National University of Singapore, 6 Science Drive 2, Singapore 117546, Singapore.}

\date{\today}

\begin{abstract}
\centerline
{\bf Abstract}
2D materials are well-known to exhibit interesting phenomena due to quantum confinement. Here, we show that quantum confinement, together with structural anisotropy, result in an electric-field-tunable Dirac cone in 2D black phosphorus. Using density functional theory calculations, we find that an electric field, $E_{\rm ext}$, applied normal to a 2D black phosphorus thin film, can reduce the direct band gap of few-layer black phosphorus, resulting in an insulator-to-metal transition at a critical field, $E_c$. Increasing $E_{\rm ext}$ beyond $E_c$ can induce a Dirac cone in the system, provided the black phosphorus film is sufficiently thin. The electric field strength can tune the position of the Dirac cone and the Dirac-Fermi velocities, the latter being similar in magnitude to that in graphene. We show that the Dirac cone arises from an anisotropic interaction term between the frontier orbitals that are spatially separated due to the applied field, on different halves of the 2D slab. When this interaction term becomes vanishingly small for thicker films, the Dirac cone can no longer be induced. Spin-orbit coupling can gap out the Dirac cone at certain electric fields; however, a further increase in field strength reduces the spin-orbit-induced gap, eventually resulting in a topological-insulator-to-Dirac-semi-metal transition.
\end{abstract}

\maketitle
\section* {Introduction}
Two-dimensional (2D) layered materials, where interlayer interactions are dominated by weak van der Waals (vdW) forces, has attracted tremendous attention in nanoelectronics~\cite{Wang}. Of particular interest are quantum confinement effects in which the electronic structure of the 2D layered material changes abruptly as a function of the number of layers. A well-known example is the indirect to direct band gap transition when 2D MoS$_2$ is reduced to a single layer~\cite{Splendiani}. Here, we show that quantum confinement results in the emergence of an electrically tunable Dirac cone in a recently discovered 2D material - 2D black phosphorus. 

Black phosphorus, a layered semiconductor material, is the thermodynamically stable form of phosphorus. Different experimental techniques, such as mechanical~\cite{Li} and liquid~\cite{Coleman} exfoliation, have been employed to thin down the bulk to a monolayer. Experimentalists have succeeded also in measuring a thickness-dependent transport gap, ranging from $\sim$0.30 eV (bulk) to 1.0 eV (monolayer)~\cite{Saptarshi}, and in fabricating field effect transistors based on ultrathin black phosphorus~\cite{Li}. Such transistors exhibit high mobility of $\sim 1000~{\rm cm^2/V\cdot s}$ and appreciably high on/off ratios, up to 10$^4$, opening up a possible potential application in nanoelectronics~\cite{Li}. In contrast to previously studied 2D materials such as graphene and MoS$_2$, 2D black phosphorus layers are highly anisotropic ( see Figure 1), with zigzag chains in one direction ({\it y} in the figure) and armchair chains in the other direction ({\it x}). This anisotropy adds another dimension of interest to 2D materials, resulting in direction-dependent optical and transport properties~\cite{Xia}.
\begin{figure}
\center
\includegraphics[width=8.0cm,clip=true]{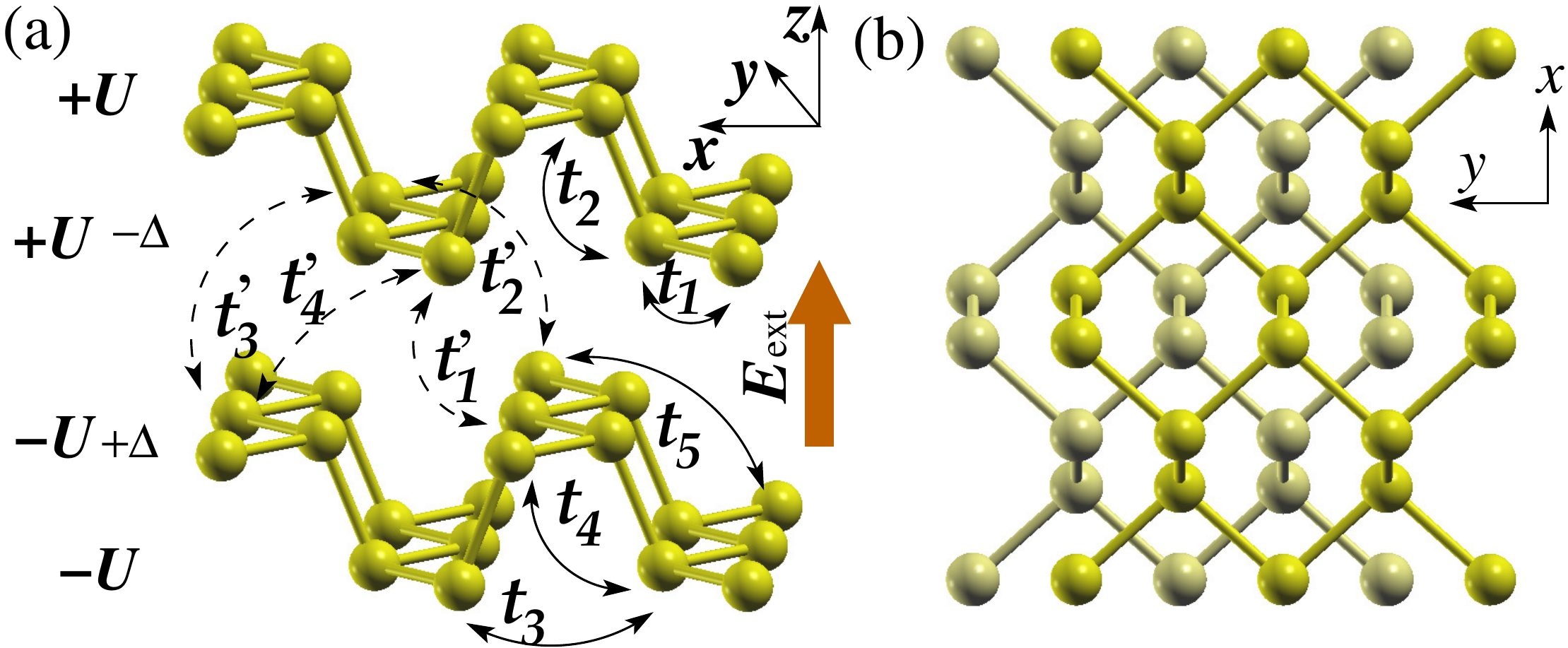}
\caption{(Color online) The optimized structure of a bilayer (2L) black phosphorus. (a) and (b) show the top and side view of 2L black phosphorus. An external vertical electric field, $E_\mathrm{ext}$ is applied along $z$-axis and  $2U$ denotes the corresponding potential difference across the layers. $\Delta$ is the potential difference between the two vertically inequivalent sites in each layer. For the sake of simplicity we use $|\Delta| = U/2$ for our tight-binding model of the bilayer system. In (a), hopping parameters of the tight-binding model are indicated. $t$ and $t'$ denote intralayer and interlayer hopping parameters respectively. In (b), dark and light yellow correspond to two different layers of black phosphorus.}
\label{fig:structure}
\end{figure}
\begin{figure*}
\centering
\includegraphics[width=16.0cm,clip=true]{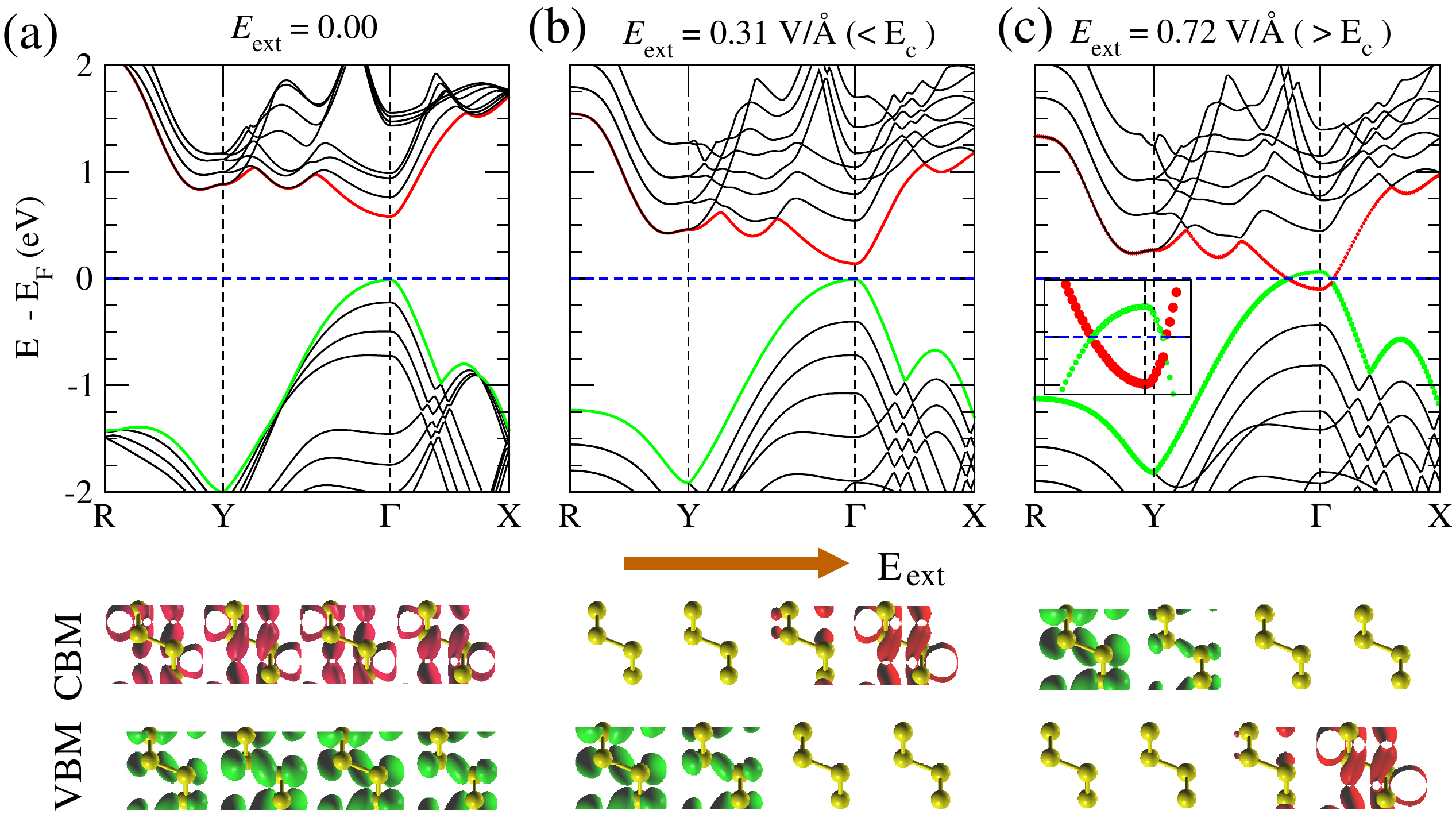}
\caption{(Color online) Electronic structure of 4L black phosphorus as a function of an external electric field, $E_\mathrm{ext}$. In (a) we present: (upper panel) the band structure and (lower panel) the charge density, respectively at the CBM (top) and VBM (bottom) for $E_\mathrm{ext}=0$. In (b) and (c), the same quantities are shown for $E_\mathrm{ext}$ = 0.31 and 0.72 V/\AA, respectively. Note that the critical field, $E_c$,  is $\sim$ 0.48 V/\AA~ for the 4L black phosphorus. In (c), where $E_{\rm ext} > E_c$, a Dirac cone appears along the $k_y$ direction, and a band gap of $\sim$39 meV opens along the $k_x$ direction. Inset in (c): zoomed-in bandstructure around the $\Gamma$ point near the Fermi level, $E_F$. Band inversion is evident in (c).}
\label{fig:band-LDOS}
\end{figure*}

In this work, we explore from first principles calculations the effect of external electric fields applied perpendicular to thin films of black phosphorus. Similar to the Stark effect previously predicted for boron nitride nanotubes, two-dimensional transition metal dichalcogenides, and other 1D and 2D materials~\cite{Ashwin, Khoo, Kapildeb, Park, Son, Gava} we find that the electric field localizes the valence band maximum (VBM) and conduction band minimum (CBM) states at opposite surfaces of the slab. The resulting potential difference between the VBM and CBM states results in a decrease in the band gap, eventually leading to an insulator-to-metal transition at a critical applied field, $E_c$. However, in contrast to other materials, we also find that for black phosphorus films below a certain critical thickness, a further increase in applied field strength results in a highly anisotropic opening in the band gap that leads to a formation of a Dirac cone. This result is not observed for black phosphorus films that are too thick. We show that the emergent Dirac cone physics arises from anisotropic interaction terms between the electric-field-induced quantum-confined VBM and CBM states in few layer black phosphorus. 

\section*{RESULTS AND DISCUSSION}
{\bf Electronic structure at zero $\bf E_{\rm ext}$} We begin our study by calculating the electronic structure of pristine black phosphorus at zero external electric field. Electronic structure calculations are performed by using density functional theory (DFT) within the Perdew-Burke-Ernzerhof parametrization (PBE)~\cite{PBE} of the generalized gradient approximation (GGA) of the exchange and correlation potential, as implemented in the {\it Quantum Espresso}~\cite{QE} package. To treat van der Waals (vdW) interactions, we employ the vdW+DF approach for the exchange-correlation functional~\cite{vdW}(See Methods for further details). 

Bulk black phosphorus has an orthorhombic crystal structure; the optimized lattice parameters of bulk black phosphorus are $a$=3.35~\AA, $b$=4.68~\AA, and $c$=11.42~\AA, in good agreement with the corresponding experimental values ($a$=3.31 \AA, $b$=4.37 {\AA}, and $c$=10.47 {\AA})~\cite{expt-lattice}. Bulk black phosphorus is a semiconductor and we predict a direct band gap of 0.43 eV at the $\Gamma $ point. Although this band gap value is slightly larger than the experimental value of 0.35 eV~\cite{Warschauer}, it is in good agreement with other vdW-corrected density functional theory calculations~\cite{Qiao}. For thin film black phosphorus, we find that the structural parameters do not change significantly. On the other hand, the band gap increases monotonically with decreasing thickness, reaching 0.99 eV for the monolayer. Interestingly, the band gap remains direct at the $\Gamma$ point when the thickness reduces from bulk to monolayer limit. 

As previously mentioned, black phosphorus is a layered compound consisting of puckered phosphorene layers with highly anisotropic bonding within each layer (Figure 1). We denote multi-layered black phosphorus by $n$L-BP, where $n$ is the number of layers. 
~\ref{fig:band-LDOS}(a) shows the electronic structure of 4L-BP, which exhibits a direct band gap of 0.60 eV at the $\Gamma$ point. The valence band maximum (VBM) and conduction band minimum (CBM) charge density are both delocalized over the layers. Our analysis of the projected density of states indicates that the charge density at the VBM and the CBM originate from phosphorus 3$s$, 3$p_x$ and 3$p_z$ orbitals (see Supplementary Figure S1). The $p_z$ orbitals have the largest contribution ($\textgreater$~70\%) to the VBM and CBM among all the orbitals.
\begin{figure}
\centering
\includegraphics[width=8.0cm,clip=true]{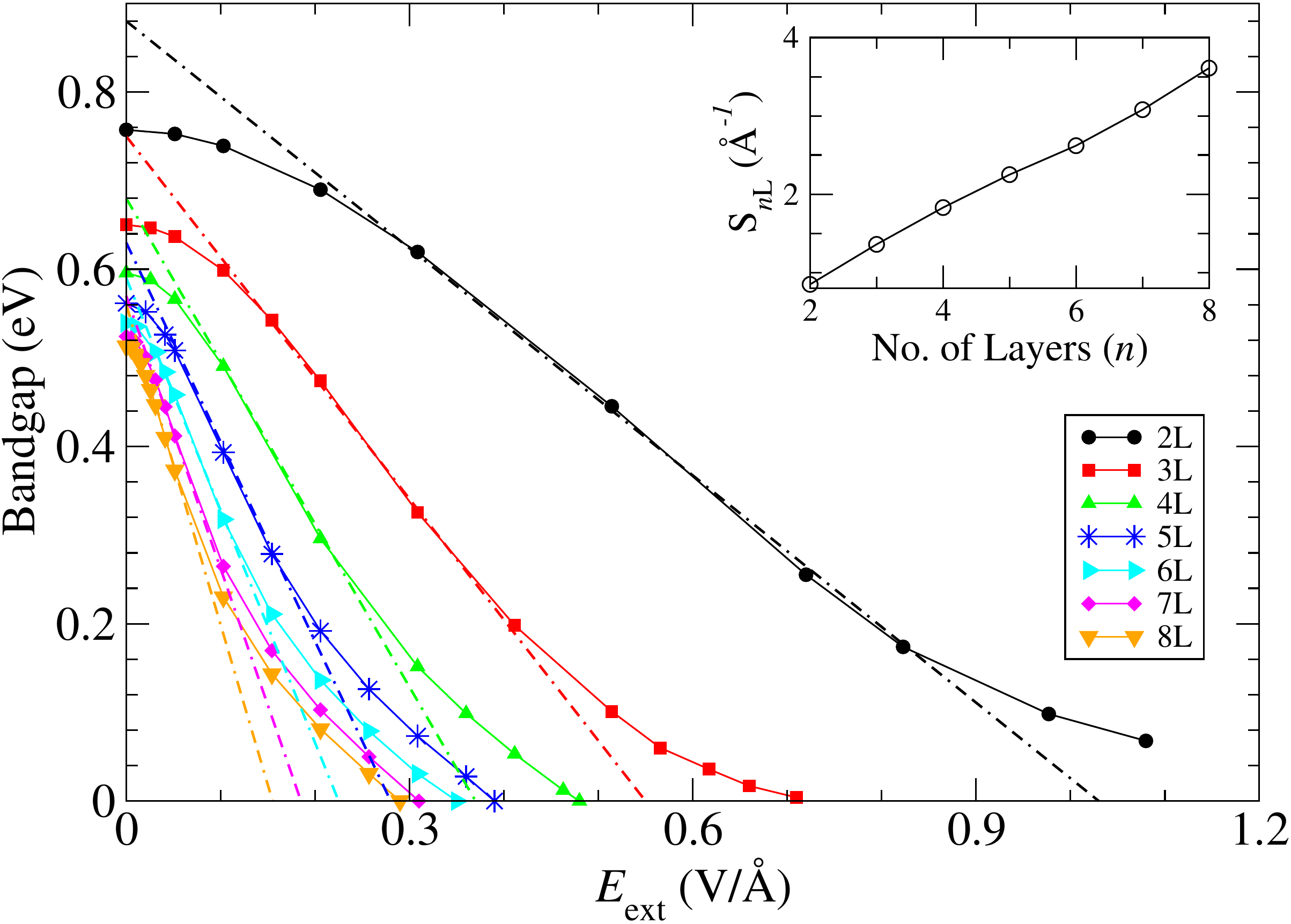}
\caption{(Color online) Variation of the (PBE+vdW) band gap with the applied electric field, $E_\mathrm{ext}$, for 2L (black circles), 3L (red squares), 4L (green triangles), 5L (blue stars), 6L (cyan triangles), 7L (magenta diamonds), and 8L (orange triangles) -BP. Inset: variation of the Stark effect coefficient, $S_{nL}$, with number of layers. Note that we consider the linear region of the band gap versus $ E_{\rm ext}$ curve to compute $S_{nL}$, and the fitted linear part is indicated by the corresponding dashed line.}
\label{fig:BGvsEext}
\end{figure}

{\bf Electronic structure for $\bf E_{\rm ext} < E_c$} We now discuss the response of the electronic structure of $n$L-BP to a static external electric field with magnitude $E_{\rm ext}$, applied normal to the layers. As $E_{\rm ext}$ increases, the band gap decreases monotonically while remaining direct at $\Gamma$, and eventually closes at a critical field, $E_c$ (see Figure~\ref{fig:band-LDOS}(b) and Figure~\ref{fig:BGvsEext}). This phenomenon essentially arises from the Stark effect: the applied electric field lifts the degeneracy among distinct layers in the thin film. After the degeneracy is lifted, states that are localized in regions with higher potential will have a higher energy due to electrostatic interaction with the external field, while those localized in regions with lower potential will have a lower energy. Consequently, the VBM becomes localized on the surface layer with higher potential, while the CBM becomes localized on the opposite surface layer with lower potential, resulting in a decrease in band gap. We note that, strictly speaking, the VBM and CBM states also have tails into the interior layers (Figure ~\ref{fig:band-LDOS}(b-c)). These tails are important for the VBM and CBM states to interact when they become energy degenerate at the critical field (see later). 

Figure~\ref{fig:BGvsEext} shows the modulation of band gap as a function of $E_{\rm ext}$ for different $n$L-BP, where $n$ runs from 2 to 8. We find that the bandgap reduces very rapidly with $E_{\rm ext}$ as the thickness gets larger. Such a thickness dependence can be easily understood by considering the potential drop across the layers. If one simply assumes that the potential drops linearly across the layers, a larger thickness requires a smaller electric field to sustain the same potential difference between the two edges. Furthermore, the band gap at zero field also decreases with increasing thickness. As such, the magnitude of the critical field, $E_c$, decreases with thickness. Indeed we observe that the value of $E_c$ drops to 0.29 eV/{\AA} for 8L-BP. In practice, the potential drop is not strictly linear in space due to screening within the layers; however, an analysis of the screening effect shows that the potential drop in the interlayer region is much larger than that in the intralayer region, and the average potential in each layer drops linearly across the layers~(see Supplementary Figure S2).

We quantify the above analysis as follows. Under the assumption that the potential drops linearly across the layers, the external field, $E_{\rm ext}$,  induces a potential difference, $\Delta V$, which can be expressed as $\Delta V = (eE_{\rm ext}/\kappa) \mean{Z}$, where $\kappa$ is the dielectric constant along the direction of the applied field and $\mean{Z}$ is defined as $\mean{Z}=\mean{Z}_{\rm V}-\mean{Z}_{\rm C}$. Here $\mean{Z}_{\rm V}$ and $\mean{Z}_{\rm C}$ denote the center positions (along the $z$-axis) of the localised VBM and CBM states, respectively~\cite{Fawei}. Since the VBM and CBM localize at opposite surfaces for large $E_{\rm ext}$, $\mean{Z} = n$c$-2\delta$, where $c$ is the layer-layer distance and $\delta$ is the position of $\mean{Z}_{\rm V, C}$ from the surfaces~\cite{Fawei}. We note that we are assuming here that $\delta$ is independent of the applied field and film thickness. Although this is not likely to be valid, deviations should be small compared to $\mean{Z}$. Under this model, we can write the change in band gap, $\Delta E_{\rm g}$, as follows:
\begin{eqnarray}\label{gap}
\Delta E_g=e\Delta V= \frac{eE_{\rm ext}}{\kappa}\mean{Z} .
\end{eqnarray}
 In general for the critical field, $E_c$, we obtain from equation~\ref{gap} that 
\begin{eqnarray}
-E_{\rm g}= \frac{eE_c}{\kappa}\mean{Z}= \frac{eE_c}{\kappa}(nc-2\delta).
\end{eqnarray}
Indeed, we find that $E_g/E_c$ versus $n$ is approximately linear (see Supplementary Figure S3), with deviations resulting from the assumption that $\delta$ is constant.

In the initial response to $E_{\rm ext}$, the  band gap varies non-linearly (likely quadratic) with $E_{\rm ext}$, and the slope of the curve vanishes at $E_{\rm ext} = 0$ (Figure 3). This is consistent with the fact that the variation in band gap should be independent of the polarity of the applied field, so that in this low field regime, the above linear analysis breaks down. However, for larger electric fields, for instance $(E_{\rm ext}$/$E_c$)\textgreater~0.1 for a specific thickness, $\Delta E_{\rm g}$ is an approximately linear function of $E_{\rm ext}$ (Figure~\ref{fig:BGvsEext}) and this linear slope can be described as 
\begin{eqnarray}\label{stark1}
\frac{dE_{\rm g}}{dE_{\rm ext}} = -eS_{n\rm L},
\end{eqnarray} where $e$ is the electron charge and $S_{n\rm L}$  is defined as the linear giant stark effect (GSE) coefficient. Combining Equation ~\ref{gap} with Equation~\ref{stark1}, the GSE coefficient can be written as
\begin{eqnarray}\label{stark2}
 S_{n\rm L}= \frac{\mean{Z}}{\kappa} = \frac{(nc-2\delta)}{\kappa}.
\end{eqnarray}
Indeed, the inset of Figure~\ref{fig:BGvsEext} shows that the Stark coefficient increases approximately linearly as a function of the number of layers.
\begin{figure*}
\centering
\includegraphics[width=17.0cm,clip=true]{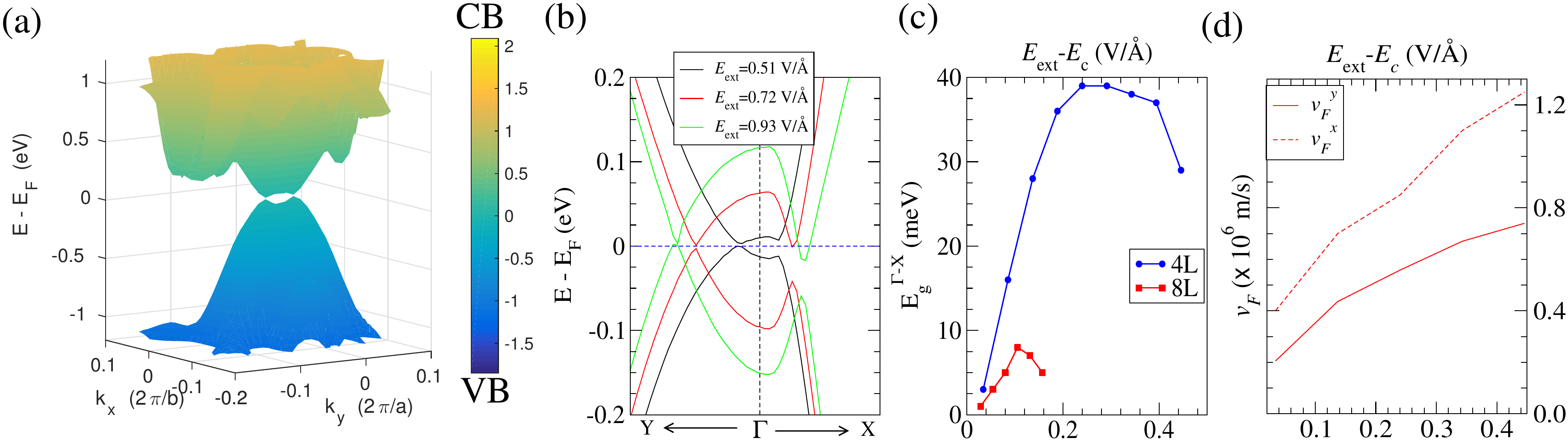}
\caption{(Color online) (a) The PBE+vdW bandstructure of 4L-BP is plotted in the 2D brillouin zone for $E_{\rm ext}$ = 0.72 V/\AA, above the critical field. The Dirac cone appears at $\pm\Lambda$. (b) The PBE+vdW bandstructure along Y-$\Gamma$-X, for the three different $E_\mathrm{ext}$ above the critical value,  namely $E_\mathrm{ext}$ = 0.51 (black), 0.72 (red), and 0.93 (green) V/\AA. (c) The bandgap that opens along $\Gamma$-X direction, $E_{g}^{\Gamma-X}$ , is plotted against ($E_{\rm ext}-E_{c}$) for 4L-BP (blue line) and 8L-BP (red line). Note that the values of $E_c$ for 4L-BP and 8L-BP are 0.48 and 0.28 V/\AA, respectively. (d) The Fermi velocity along $k_x$, $v_F^x$ (red dashed line), and along $k_y$, $v_F^y$ (red solid line) are plotted as a function of ($E_{\rm ext}-E_{c}$) for 4L-BP. $v_{\rm F}$ is calculated as $v_{\rm F}=\delta E/\delta k$ at the Dirac point.}
\label{fig:diraccone}
\end{figure*}

At the critical field, the system becomes metallic when the VBM and CBM touch at $\Gamma$. Unlike the electrically driven Stoner magnetism previously reported for MoS$_2$ nanoribbons~\cite{Kapildeb}, there is no spin polarization in this system (Supplementary Figure S4). This is because the density of states at the Fermi level is much smaller in the 2D system than in the 1D nanoribbons.

{\bf Electronic structure beyond critical field} An interesting and unique effect is observed in very thin-layer BPs (for example in 4L-BP), when an electric field larger than the critical field is applied. Specifically, a Dirac cone appears at a $k$-point along $\Gamma$-Y, which we shall call $\Lambda$, as shown in Figure~\ref{fig:band-LDOS}(c) and Figure~\ref{fig:diraccone}(a-b). The valence band manifold and conduction band manifold touch at exactly two points, $\pm\Lambda$, in the 2D Brillouin Zone [Figure~\ref{fig:diraccone}(a)]. By plotting the charge density of the highest occupied and lowest unoccupied states at k-points close to the Dirac point(see Supplementary Figure S5), we see that the bands are inverted at the Dirac point, indicating that we have a new topological phase. As black phosphorus films exhibit a direct band gap even in the presence of the external electric fields, the band inversion in itself may not be too surprising. However, the fact that the valence and conduction band manifolds touch at only one point, the Dirac point, is intriguing and deserves further analysis.

As $E_{\rm ext}$ increases beyond the critical value, the Dirac point, $\Lambda$, shifts progressively from $\Gamma$ toward Y in the 2D Brillouin zone [Figure~\ref{fig:diraccone}(b)]. This shifts occurs simultaneously with an enhancement of the corresponding Dirac Fermi velocity, $v_{\rm F}$~[Figure~\ref{fig:diraccone}(c)]. The existence of the Dirac point is associated with an asymmetry between the $\Gamma$-Y and $\Gamma$-X directions - so that the valence and conduction bands touch only at the Dirac points and not in a ring in the 2D Brillouin Zone; neither does a band gap open up throughout the 2D Brillouin Zone. We quantify this asymmetry and therefore also the robustness of the Dirac electronic structure by the band gap along $\Gamma$-X, $E_{\rm g}^{\Gamma-X}$. As shown in Figure~\ref{fig:diraccone}(c), $E_{\rm g}^{\Gamma-X}$ increases initially as $E_{\rm ext}$ increases beyond the critical field; however, as $E_{\rm ext}$ continues to increase, $E_{\rm g}^{\Gamma-X}$ decreases. For 4L-BP, $E_{\rm g}^{\Gamma-X}$ reaches a maximum value of $\sim$39 meV at $E_{\rm ext}$=0.72 V/\AA~, where the corresponding $x~{\rm and}~y$-components of $v_F$ are 0.84$\times~10^{6}$ and 0.56$\times 10^{6}$ m/s, respectively. These Fermi velocities are as large as the predicted value of $\sim 0.86\times10^6$ m/s in graphene~\cite{Daniel}. Interestingly, we find that the maximum value of $E_{\rm g}^{\Gamma-X}$  reduces to 8 meV when the thickness of black phosphorus increases to 8L [Figure~\ref{fig:diraccone}(c)], and vanishes for the 12L case~(see Supplementary Figure S6).  Salient features that arise from the above analysis are that $E_{\rm g}^{\Gamma-X}$ cannot be arbitrarily increased with electric field strength, and that $E_{\rm g}^{\Gamma-X}$ is smaller for thicker black phosphorus films.
\begin{figure}[htbp]
\centering
\includegraphics[width=8.0cm,clip=true]{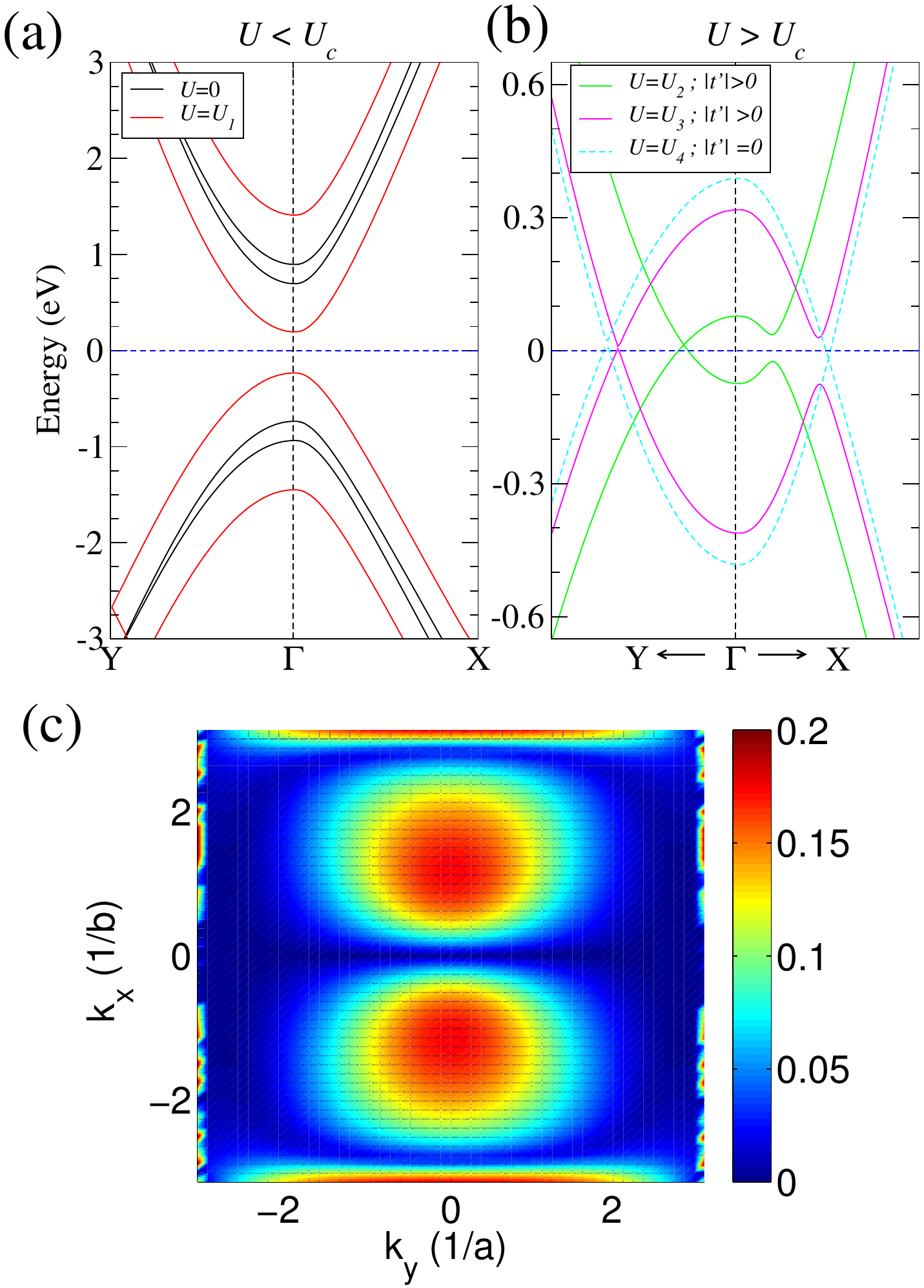}
\caption{(Color online) The bandstructure of bilayer black phosphorus calculated by using a simple tight-binding model. In (a) the bandstructure is plotted for $E_\mathrm{ext}$ = 0 (black line) and $E_\mathrm{ext}$ \textless $E_c$ (red line). In (b) the bandstructures are plotted for three different $E_\mathrm{ext}$ above the critical value, where $U_1 $\textless $U_2$ \textless $U_3$ \textless $U_4$.  For the applied bias $U_2$ (green lines) and $U_3$ (magenta lines), the interlayer hopping, $t'$, is finite and kept fixed, whereas $t'=0$ for the applied bias $U_4$ (cyan dashed lines). The horizontal dotted line corresponds to the Fermi level. (c) The $k_x-k_y$ plot of the magnitude of $<\phi_{\rm V}|H(t')|\phi_{\rm C}>$, where $\phi_{\rm V}$ and $\phi_{C}$ are the eigenvectors of the Hamiltonian $H$ at $t'=0$ for the highest occupied and  the lowest unoccupied energy  eigenvalues, respectively. We use $t'$=0.2 eV for the 2D plot. }
\label{fig:TB}
\end{figure}
To understand the above intriguing phenomena, we develop a simple tight-binding model that captures the essential physics of black phosphorus thin films in the presence of an external electric field.  

{\bf Tight-binding model}
For simplicity we consider a tight-binding model for BP films that are two-layer thick (bilayer BP). Since we are concerned with the valence band (VB) and conduction band (CB) which stem from $p_z$ orbitals, we include only $p_z$ orbitals in our model. In the presence of large external electric fields, the VB and CB are mainly localized at the opposite surface layers of the thin film; therefore, the bilayer model can effectively also represent the VB and CB electronic structure in thicker films.

We begin with a standard tight-binding Hamiltonian:
\begin{eqnarray}
H = \sum_i  (\varepsilon_i+U_i)~c_i c_{i}^{\dagger} + \sum_{<i,j>}t_{ij}~ c_ic_j^{\dagger} + \sum_{<i,j>} t'_{ij}~c_ic_j^{\dagger}
\end{eqnarray}
where $\varepsilon_i$ is the energy of $i$-th site, $U_i$ is the potential shift at the $i$-th site upon the applied vertical field (see figure~\ref{fig:structure}), $c_{i}^{\dagger}$($c_j$) is the creation (annihilation) operator of electrons at site $i(j)$, and $t_{ij}$ and $t'_{ij}$ are respectively the intralayer and interlayer inter-site hopping integrals as depicted in figure~\ref{fig:structure}. In the first instance, we include hopping terms up to the fourth and fifth nearest neighbor for the interlayer and intralayer interactions respectively, using the parameters reported previously in Ref.~\cite{Rudenko}. This provides an accurate description of the VB and CB, as given by the $GW$ approximation. Figure~\ref{fig:TB}(a) shows that our model nicely reproduces the band dispersion of the VB and CB in comparison with that of the $GW$ bands~\cite{Rudenko}.

When an external field is applied, the on-site energy of each site, $\varepsilon_i$, is rigidly shifted by the corresponding potential difference, $U_i$ (see Figure~\ref{fig:structure}). For small $U_i$ [ Figure~\ref{fig:TB}(a)] we see that the band gap closes as described above. When $U_i$ increases above a critical value, $U_c$, the Dirac cone emerges along the $\Gamma$-Y direction, just as was predicted by DFT, with the position of the Dirac cone shifting with the size of $U_c$ (Figure~\ref{fig:TB}(b), green and magenta curves).  

However, interestingly, when the interlayer interaction terms $t'$ are set to zero, $U > U_c$ causes the band gap to vanish all over the two-dimensional Brillouin Zone~(Figure~\ref{fig:TB}(b), cyan curve). This finding indicates that the interlayer interaction is critical to the emergence of a Dirac cone. Intuitively, we expect the VB and CB states to interact via these interlayer terms, since the states are localized at opposite surfaces of the thin film. The opening of a band gap everywhere in the Brillouin Zone except along $\Gamma$-Y thus results from interactions between the VB and CB states, mediated by the interlayer hopping integrals. This also explains why the maximum value of $E_{\rm g}^{\Gamma-X}$ decreases as the thickness of the BP film increases and the interactions between the VB and CB become smaller. Thus, while the critical field $E_c$ may become smaller with thicker films due to a decrease in the band gap with increasing thickness, the Dirac cone phenomenon will be absent beyond a critical thickness even with gap closure. We find that within DFT, when the film thickness increases to 12L, $E_{\rm g}^{\Gamma-X}$ becomes negligible (equal to the precision of our calculations)(see Supplementary Figure S6). Quantum confinement is therefore important for the Dirac cone phenomenon. 
\begin{figure}
\centering
\includegraphics[width=8.0cm,clip=true]{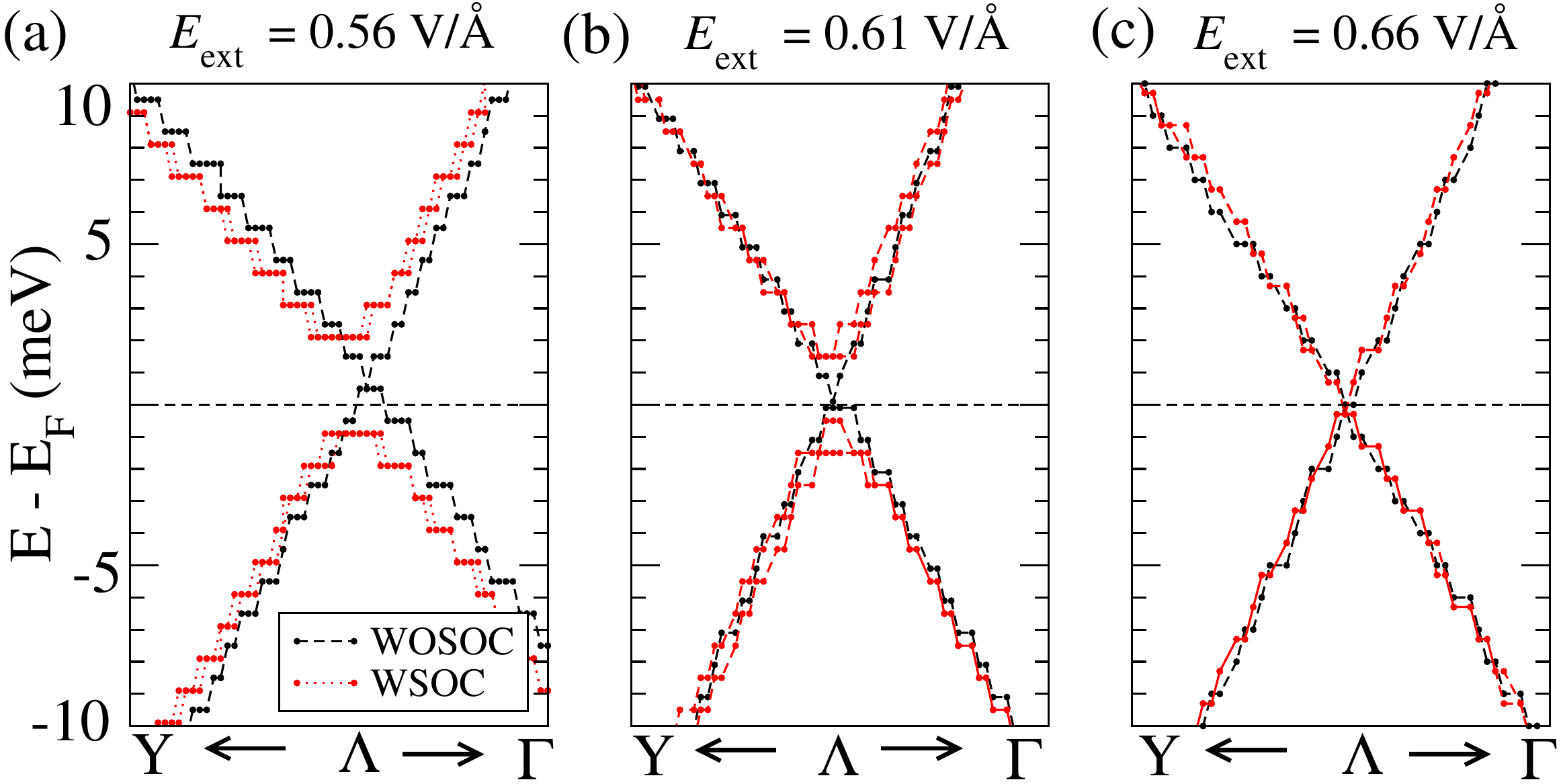}
\caption{(Color online) Effect of spin-orbit coupling on the band structure of 4L black phosphorus, zoomed-in at the $\Lambda$ point. Black and red lines denote the bandstructure calculated without and with spin-orbit coupling, respectively, at $E_{\rm ext}$ = (a) 0.56, (b) 0.61 and (c) 0.66 V/\AA.}
\label{fig:WSOC}
\end{figure}

The unique Dirac cone band structure implies also that for $U > U_c$, there is no interaction between the VB and CB states along the $\Gamma$-Y direction, but non-zero interactions along all other directions. This is a direct consequence of the anisotropic interlayer interactions in black phosphorus.  Indeed, as shown in Figure~\ref{fig:TB}(c), the $k$-dependent Hamiltonian overlap matrix elements between the VB and CB states are highly anisotropic, going to zero along $k_x = 0$, i.e. along the $\Gamma$-Y direction, but non-zero elsewhere in the Brillouin Zone. 

From Figure~\ref{fig:TB}(c), we can also see that for a fixed electric field strength, the interaction between the VB and CB states increases as we move along $\Gamma$-X away from the $\Gamma$ point. As the electric field strength is increased, the VB and CB cross further away from $\Gamma$. This explains why $E_{\rm g}^{\Gamma-X}$ initially increases as $E_{\rm ext}$ is increased [ Figure~\ref{fig:diraccone}(c)]. However, a larger electric field strength also implies greater localization of the VB and CB states, and therefore a smaller overall interaction term.  These two effects compete with each other, resulting in an initial increase and then a subsequent decrease of $E_{\rm g}^{\Gamma-X}$ as  $E_{\rm ext}$ increases [ Figure~\ref{fig:diraccone}(c)].

Thus, we have shown that the Dirac cone phenomenon arising for $E_{\rm ext} > E_c$ stems from a combination of quantum confinement and anisotropy effects in black phosphorus.
\begin{figure}
\centering
\includegraphics[width=8.0cm,clip=true]{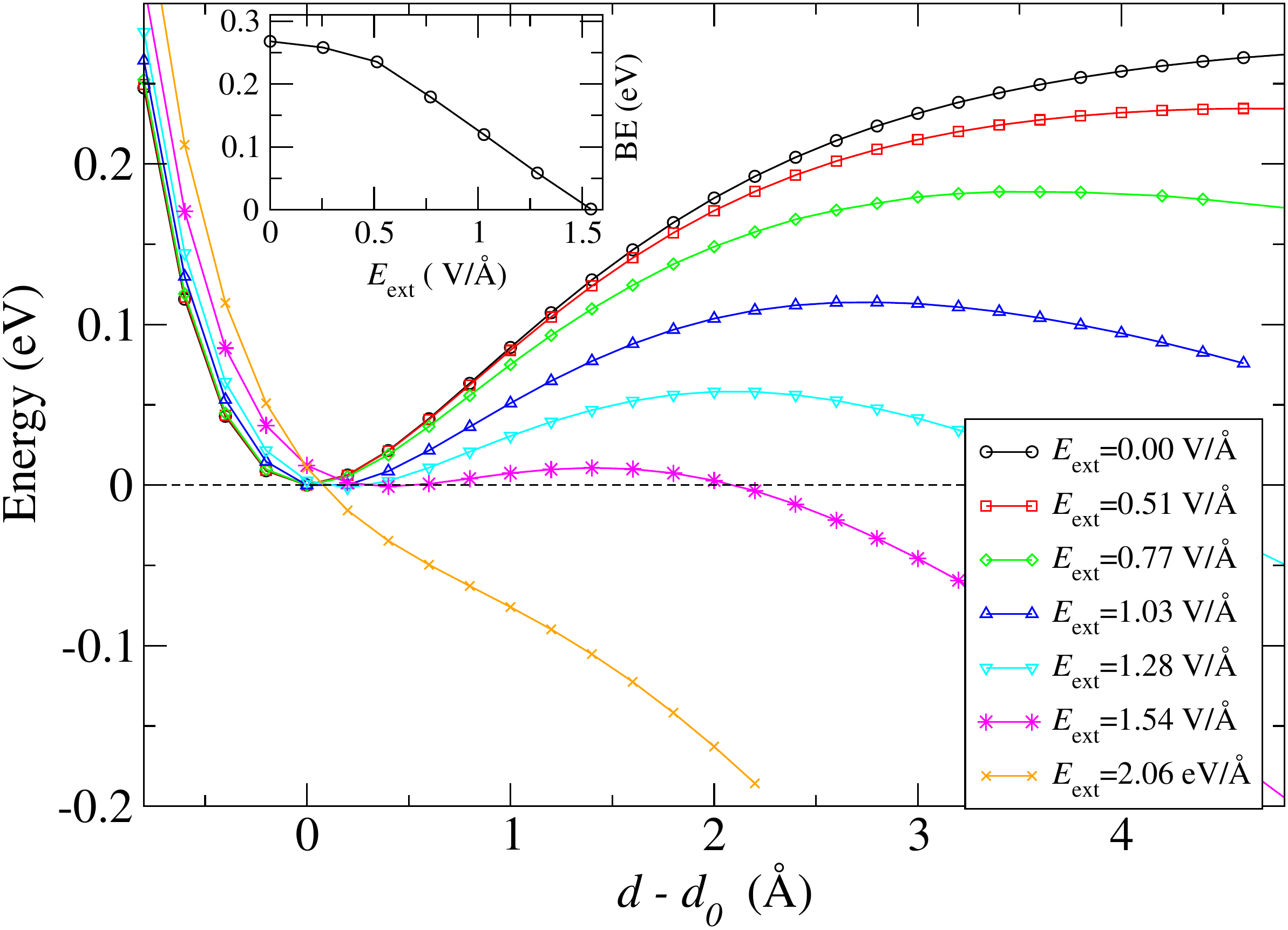}
\caption{(Color online) Total energy per unit cell plotted against $d-d_0$ for different values of $E_\mathrm{ext}$ in V/\AA. $d$ is the inter-layer distance and $d_0$=3.59~\AA is the equilibrium inter-layer distance. Inset shows the variation of binding-energy (BE), calculated within the PBE+vdW functional, as a function of the applied electric field, $E_\mathrm{ext}$, for 2L black phosphorus.}
\label{fig:binding}
\end{figure}

{\bf Bandstructure with spin-orbit correction} 
As already mentioned before, the band gap remains direct at the $\Gamma$ point until the the gap closes completely. As a consequence, the band gap does not change significantly by the inclusion of spin-orbit coupling (SOC). However, the role of  SOC becomes preeminent when $E_{\rm ext} > E_c$. Interestingly, Figure~\ref{fig:WSOC}(a) shows that the Dirac cone is gapped out in the band structure by including the SOC, which demonstrates that the system is likely to be a quantum spin Hall insulator. The prediction can be explicitly confirmed by calculating the $Z_2$ topological invariant of the system~\cite{TI-zahid}, as shown in Ref ~\cite{Zunger}. For 4L-BP, the maximum SOC-induced gap-opening at the Dirac point is as large as 3 meV at $E_{\rm ext}$= 0.56 V/\AA. When $E_{\rm ext}$ is further increased, the SOC-induced gap decreases (to 2meV at $E_{\rm ext}$= 0.61 V/\AA) and subsequently, the Dirac semimetal phase appears at $E_{\rm ext}$=0.66 V/\AA~along the $\Gamma$-Y direction (Figure~\ref{fig:WSOC}). Thus, the external electric field can be used to tune the opening and closure of the gap along $\Gamma$-Y when SOC is included. It is interesting to mention that the band inversion is controlled by electrically driven band overlap between the VB and CB, unlike other topological insulators, for example Bi$_2$Se$_3$, where band inversion is caused by spin-orbit coupling.

{\bf Robustness of results against functional}
Our results above are robust against the exact choice of functional. Specifically, with the PBE functional, the band gap also decreases monotonically with thickness (see supplementary figure S7), but the rate of reduction of the PBE-bandgap is different from that of the vdW-corrected one. However, the qualitative physics, i.e. both the thickness-dependent bandgap, and the external field induced bandgap-closure and Dirac cone, consistently appear regardless of the functional. Notably, the strength of critical field required for the insulator-to-metal transition of bilayer BP is almost the same for both PBE and PBE+vdW functionals. Moreover, although many-electron $GW$ calculations are in principle required for accurate calculation of the band gap, the PBE and vdW corrected functionals underestimate and overestimate the bulk band gap respectively, and in both cases, the same physics emerges.
\begin{figure}
\centering
\includegraphics[width=8.0cm,clip=true]{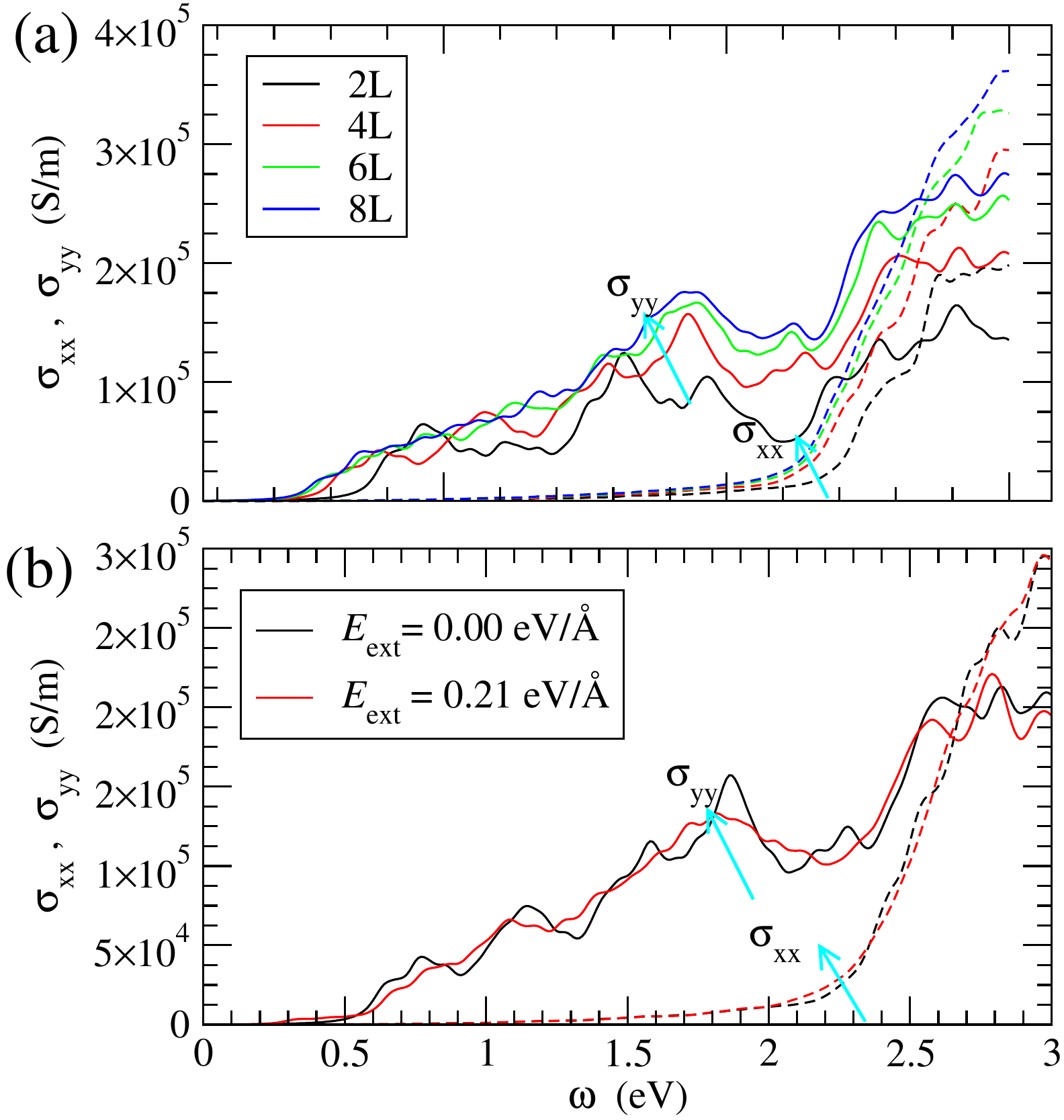}
\caption{(Color online) Tuning of optical properties by thickness and applied $E_\mathrm{ext}$. (a) The optical conductivity of 2L (black), 4L (red), 6L (green), and 8L (blue) BP, plotted as a function of frequency. Solid and dashed lines denote two different components of optical conductivity, $\sigma_{xx}$ and $\sigma_{yy}$, respectively. (b) $\sigma_{xx}$ (solid lines) and $\sigma_{yy}$ (dashed lines) of 4L-BP are plotted against the frequency for two different applied fields, $E_\mathrm{ext}$ = 0.0  (black) and 0.21 (red) V/\AA.}
\label{fig:optical}
\end{figure}

{\bf Binding energy}
We also calculate the total energy as a function of interlayer distance in bilayer BP for different $E_{\rm ext}$, as shown in Figure~\ref{fig:binding}. We find that the binding energy reduces by increasing the strength of $E_{\rm ext}$ and eventually vanishes for a field of $E_{\rm ext} \sim$ 1.6 V/\AA, at which point the layers becomes unbounded. This field is much larger than the critical fields at which the Dirac cone phenomenon was predicted. Notably the equilibrium distance between the layers is almost independent of $E_{\rm ext}$. This result suggests that an external field can trigger the exfoliation process of black phosphorus as an alternative to mechanical~\cite{Li} and chemical exfoliation~\cite{Coleman}. Note that few-layer graphene has been successfully exfoliated in the presence of $E_{\rm ext}$~\cite{Liang}.

{\bf Optical Properties}
Given the predicted changes in band gap as a function of thickness and electric field strength, we also consider a related quantity - the optical conductivity of the system. Figure~\ref{fig:optical}(a) shows both the $xx$ and $yy$-components of the optical conductivity, $\sigma_{xx}$ and $\sigma_{yy}$, as a function of frequency for 2L, 4L, 6L and 8L structures. We find that the onset of $\sigma_{xx}$ and $\sigma_{yy}$ depend on the dipole transitions at a point near R, and at the $\Gamma$ point, respectively. As the bandgap reduces with increasing thickness, the onset of $\sigma_{yy}$ also decreases, while that of $\sigma_{xx}$ decreases slightly. By applying an external field to 4L-BP, we find that the electric-field induced reduction in direct bandgap decreases the onset of $\sigma_{yy}$ whereas that of $\sigma_{xx}$ remains almost unchanged [ Figure~\ref{fig:optical}(b)].  

\section*{CONCLUSIONS}
In summary, we have shown using first principles calculations that the application of an external electric field to few-layer BP results in a reduction in the direct band gap of the material, eventually resulting in an insulator-to-metal transition at a critical field. Above this critical field, a Dirac cone emerges at $\Lambda$ along the $\Gamma-Y$ direction, provided that the BP film is sufficiently thin. We quantify the robustness of the Dirac cone by the energy gap in the $\Gamma-X$ direction, $E_{\rm g}^{\Gamma-X}$. Using a tight-binding model, we show that quantum confinement and structural anisotropy are critical for the emergent Dirac cone phenomenon. These conclusions further explain the subtle dependence of $E_{\rm g}^{\Gamma-X}$ on the electric field strength and film thickness, providing us with a handle to tune the resulting electronic structure. The electric field strength can be used to tune the position of $\Lambda$ and the Dirac-Fermi velocities, the latter being similar in magnitude to those in graphene. We also show that by including spin-orbit coupling, the Dirac point can be gapped out at certain electric fields, but increasing the electric field strength further can change the material from a topological insulator to a Dirac semi-metal. We have thus shown that the electronic band structure of few-layer black phosphorus can be tuned in very interesting ways by an applied electric field. These results will motivate the search for other direct band gap materials where quantum confinement and structural anisotropy may lead to tunable emergent Dirac cone physics.

\section*{METHODS}
Electronic structure calculations are performed by using density functional theory within the generalized gradient approximation (GGA) of the exchange and correlation potential Perdew-Burke-Ernzerhof parametrization (PBE)~\cite{PBE} as implemented in the {\it Quantum Espresso}~\cite{QE} package. To treat van der Waals (vdW) interactions, we employ the vdW+DF approach for the exchange-correlation functional~\cite{vdW}. The wavefunction and the charge density are expanded using energy cutoffs of 50 and 300 Ry, respectively.  Brillouin zone sampling is done by using a (10$\times$14$\times$1) Monkhorst-Pack $k$-grid. Periodic boundary conditions have been included and a vacuum layer of at least 20~{\AA} is included to suppress the interaction between the periodic images. The conjugate gradient algorithm is used to obtain optimized geometries, where all the atoms in the unit cell are allowed to relax until the forces on each atom are less than 0.002 eV/{\AA}, and the coordinates are fixed in the calculations using the finite external electric field. We have checked that for 4L-BP, the results are qualitatively similar when the atoms are relaxed in the presence of the applied electric field~(see Supplementary Figure S8).

Simulations with an external electric field are performed using a periodic sawtooth-type potential perpendicular to the layers~\cite{swatooth}. In order to remove the effect of electrostatic interactions  between the periodic supercells,  a dipole correction scheme suggested by L. Bengtsson~\cite{dipole}, is included.

The real part of the optical conductivity is computed, using the following formula: $\Re(\sigma) = \omega \times \Im (\epsilon)$. Here the imaginary part of the dielectric tensor, $\Im (\epsilon_{\alpha \beta}(\omega))$, can be expressed as
\begin{eqnarray}
 \Im (\epsilon_{\alpha \beta}(\omega))=\frac{4 \pi e^2}{\Omega N_k  m^2}\sum_{v,c}\int_{BZ}
\frac{\hat{\textbf{M}}_{\alpha \beta}\times f(E_{\textbf{k},v})}{\nabla_k(E_{v}-E_{c})}... \nonumber \\
...\bigg[\delta(E_{\textbf{k},c}-E_{\textbf{k},v}+\hbar\omega)+\delta(E_{\textbf{k},c}-E_{\textbf{k},v}-\hbar\omega)\bigg],
\end{eqnarray}
where $e~{\rm and}~m$ are the electronic charge and mass, respectively, $\Omega$ is the lattice volume, $N_k$ is the total number of k-points, the indices $c$ and $v$ denote the conduction and the valance band, respectively, and $f$ is the Fermi-Dirac distribution. $\hat{\textbf{M}}_{\alpha\beta}$ is defined as follows:~$\hat{\textbf{M}}_{\alpha \beta}=\langle u_{\textbf{k},c}\vert\hat{\textbf{p}}_{\alpha}\vert u_{\textbf{k},v}\rangle
\langle u_{\textbf{k},v}\vert\hat{\textbf{p}}_{\beta}^{\dagger}\vert u_{\textbf{k},c}\rangle$, where $\vert u_{\bf k}\rangle$ and $\hat{\bf P}$  are the single particle Bloch function and the dipole moment operator, respectively.  An uniform k mesh of 36$\times$26$\times$1 is used for self consistent steps in order to calculate the optical conductivity. Note that our calculation does not include many-body effects which are required to obtain an accurate description of the optical conductivity.

\section*{ACKNOWLEDGMENTS} We gratefully acknowledge support from the National Research Foundation, Singapore, for funding under the NRF Fellowship (NRF-NRFF2013-07) and under the NRF medium-sized centre programme. We thank the Graphene Research Center for the computational resources provided. \\

\section*{Author contribution}
KD performed the calculations. KD and SYQ planned the study. SYQ provided supervision. KD and SYQ wrote the manuscript.

\section*{Additional Information}
The authors declare no competing financial interests.


{\Large \bf Supplementary Information for "Quantum-confinement and Structural Anisotropy result in Electrically-Tunable Dirac Cone in Few-layer Black Phosphorous"} \\
 



\begin{figure*}[htbp]
\centering
\includegraphics[width=17.0cm,clip=true]{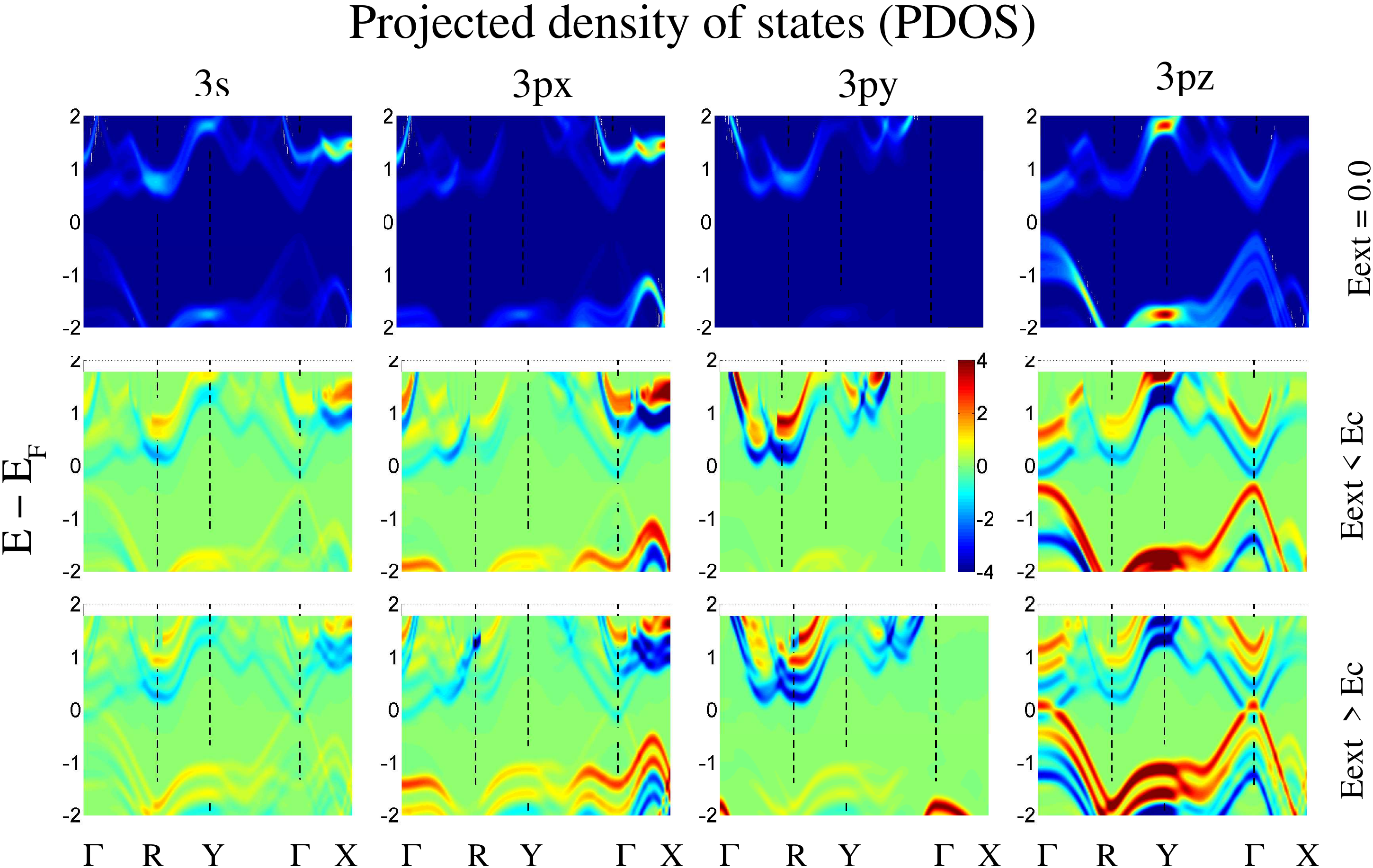}
\caption{{\bf Supporting Figure S1:} The band-decomposed projected density of states (PDOS) on different orbitals for 4 layer black phosphorus. The densities of states are projected on 3$s$, 3$p_x$, 3$p_y$ and 3$p_z$  orbitals. Upper, middle and lower panels show the PDOS at $E_{\rm ext}$ = 0.00, 0.21, and 0.72 V/\AA, respectively. Note that the critical field is $E_c$ = 0.48 V/\AA. At $E_{\rm ext}$ = 0, the PDOS is summed over all the atoms in the unit-cell. For finite fields, for visualization purposes, we use the negative colour scale for PDOS that comes from the two layers kept at positive potential, and the positive colour scale for PDOS coming from the other two layers kept at negative potential, as indicated in Figure 1 in the main text.}
\label{S1}
\end{figure*}

\begin{figure*}[htbp]
\centering
\includegraphics[width=16.0cm,clip=true]{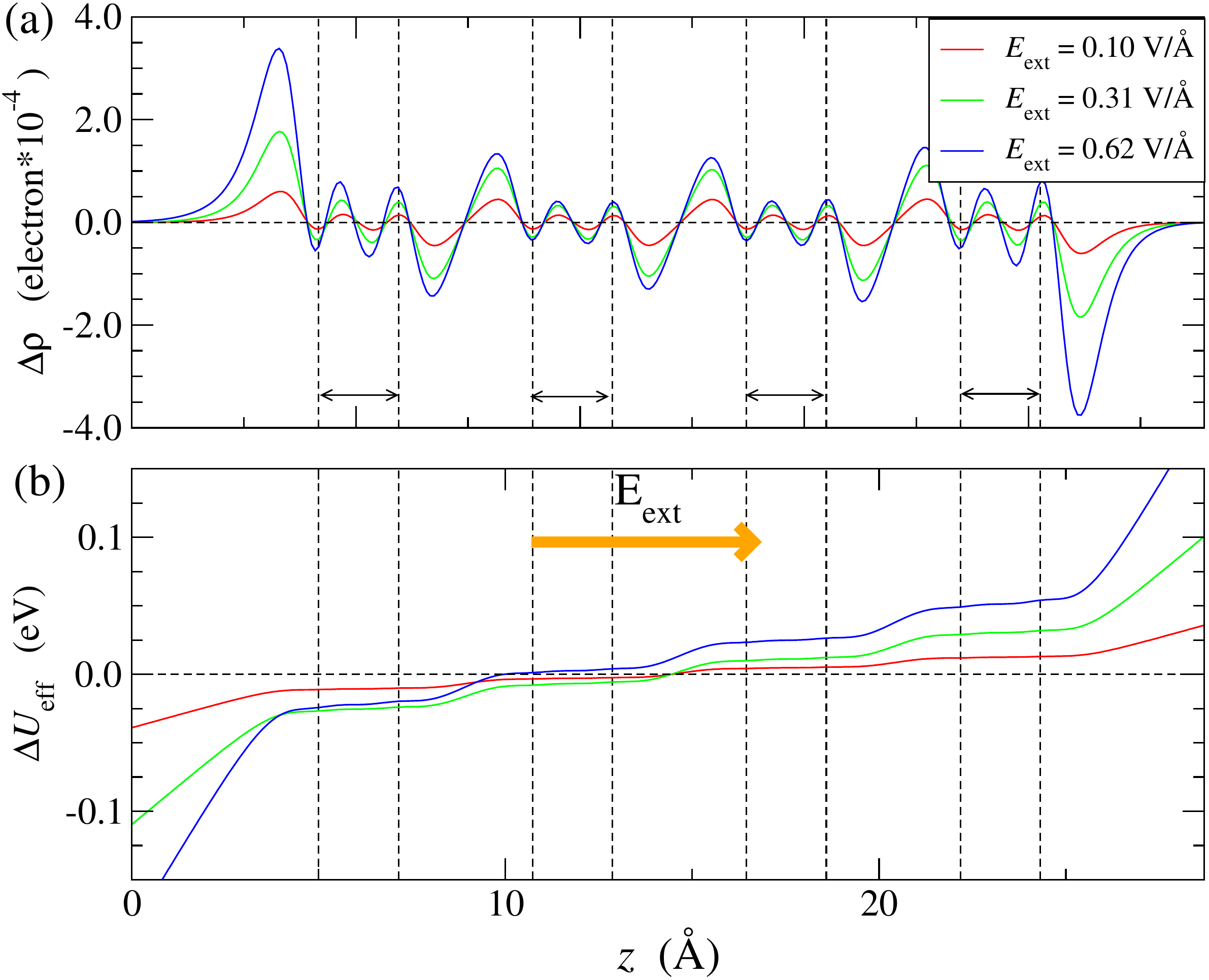}
\caption{{\bf Supporting Figure S2:} (a) Electric field induced charge density distribution across the layers, $\Delta\rho=\rho(E_{\rm ext})-\rho(E_{\rm ext}=0)$, for 4L-BP and different values of $E_{\rm ext}$. $\rho(E_{\rm ext})$ is the charge density for an external field, $E_{\rm ext}$. $\rho(E_{\rm ext})$ is averaged over the transverse direction ($xy$-plane) and plotted along the out-of-plane direction ($z$-axis). The dashed vertical lines denote the positions of each atomic plane in 4L-BP. (b) Effective electrostatic potential across the layers, $\Delta U_{\rm eff} = U(E_{\rm ext})-U(E_{\rm ext}=0)$, for 4L-BP and different values of $E_{\rm ext}$. }
\label{S2}
\end{figure*}

\begin{figure*}[htbp]
\centering
\includegraphics[width=12.0cm,clip=true]{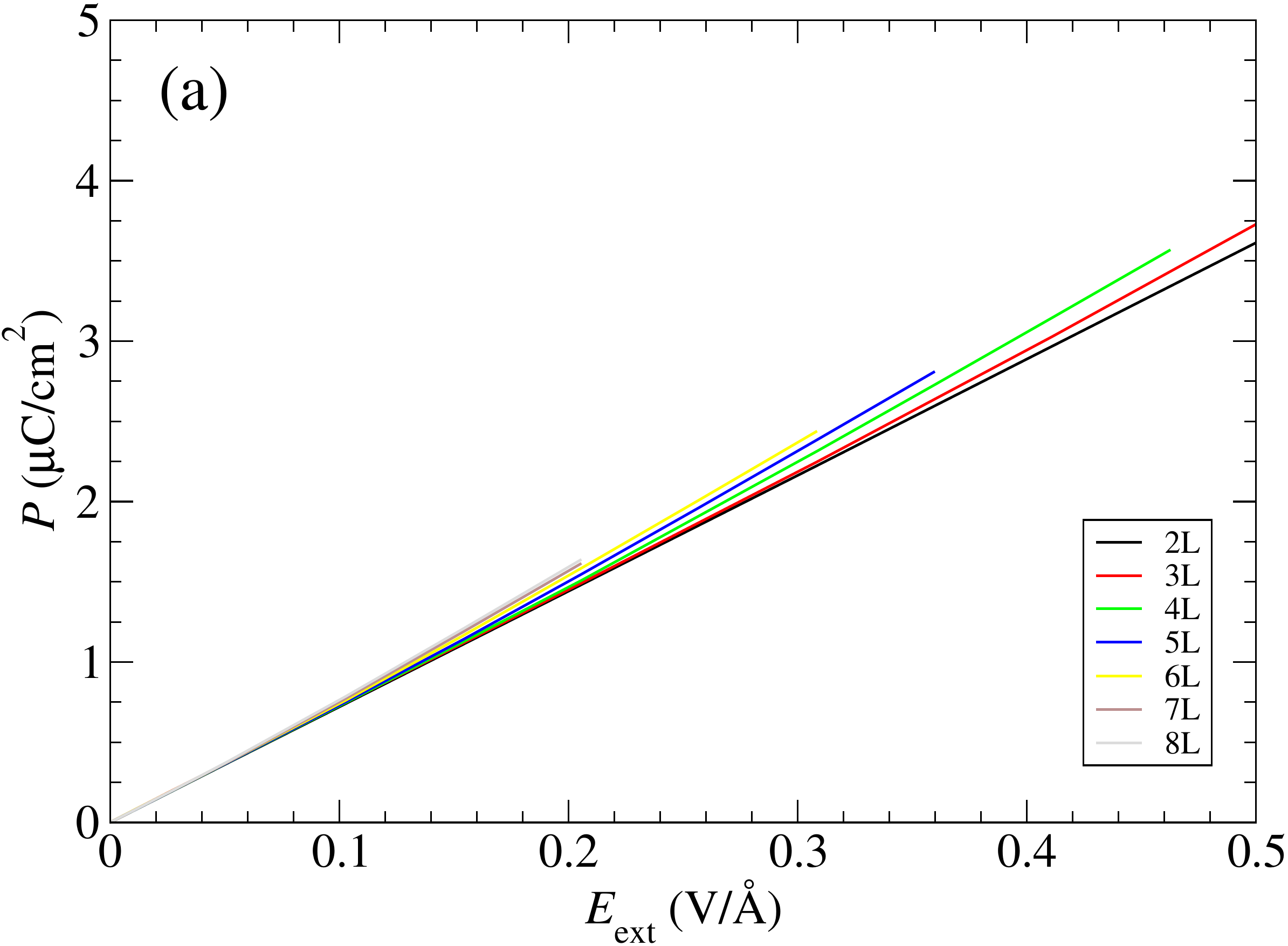}
\includegraphics[width=12.0cm,clip=true]{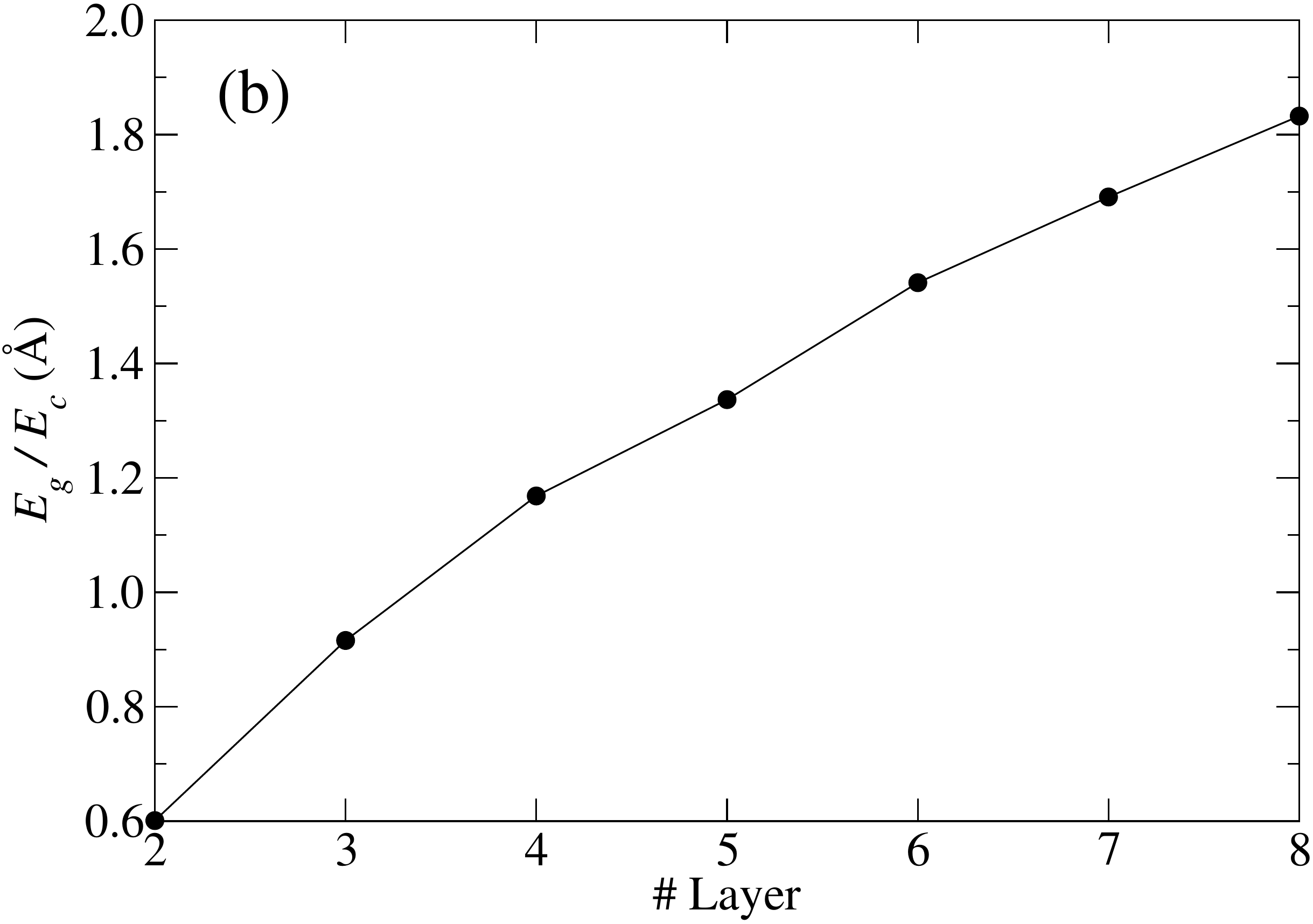}
\caption{{\bf Supporting Figure S3:} (a) The polarization (dipole moment per unit area) , $P$, is plotted against the applied external field strength for 2L (black), 3L (red), 4L (green), 5L (blue), 6L (yellow), 7L (brown) and 8L (grey)-BP. (b)  ($E_{g}/E_{c}$ is plotted as a function of the number of layers.}
\label{S3}
\end{figure*}

\begin{figure*}[htbp]
\centering
\includegraphics[width=16.0cm,clip=true]{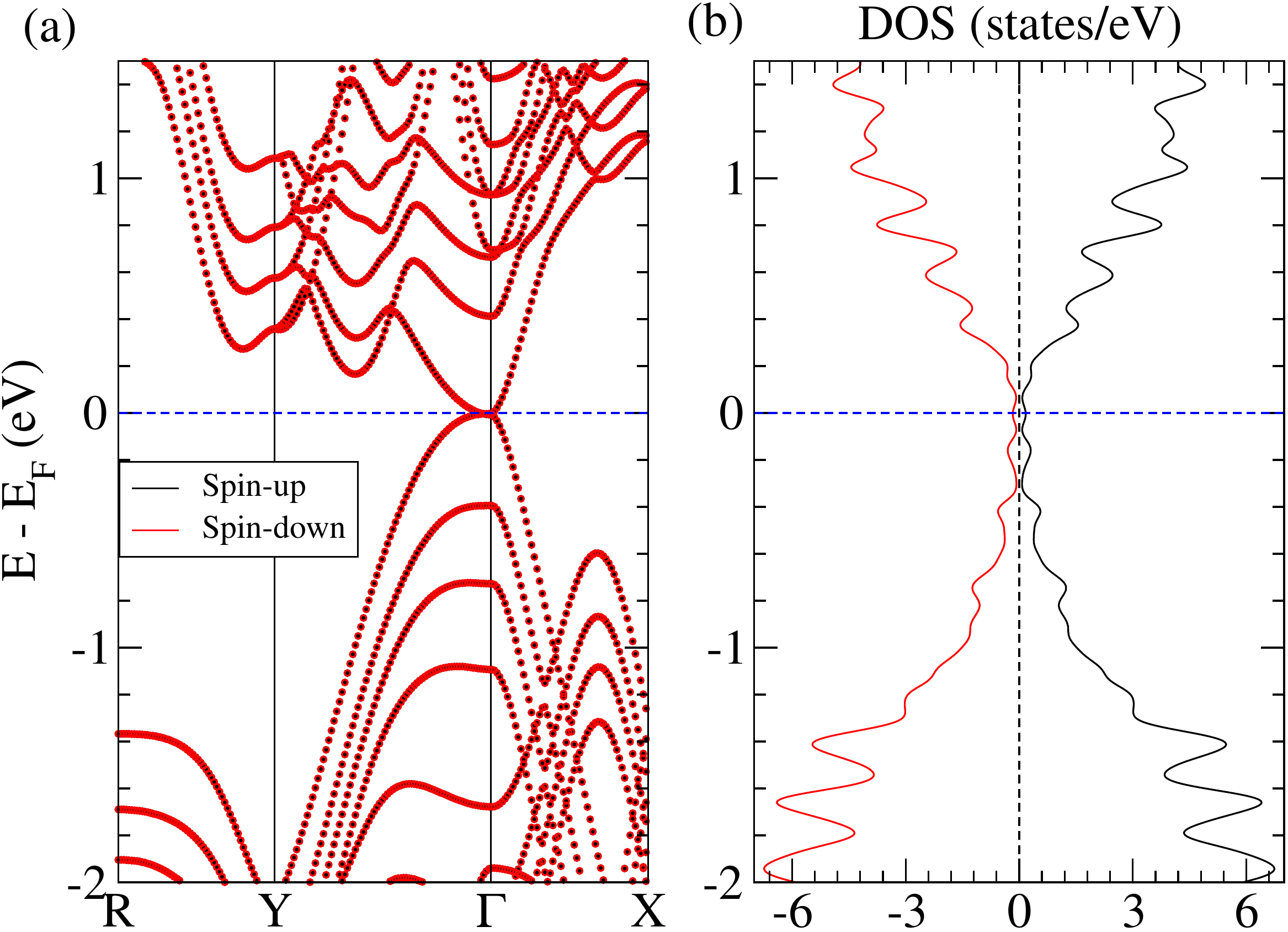}
\caption{{\bf Supporting Figure S4:} (a) Spin-polarized bandstructure and (b) density of states of 4L-BP. The blue dashed line denotes the Fermi level. Red and black lines denote the spin-up and spin-down component of the DOS, respectively.}
\label{S4}
\end{figure*} 

\begin{figure*}[htbp]
\centering
\includegraphics[width=16.0cm,clip=true]{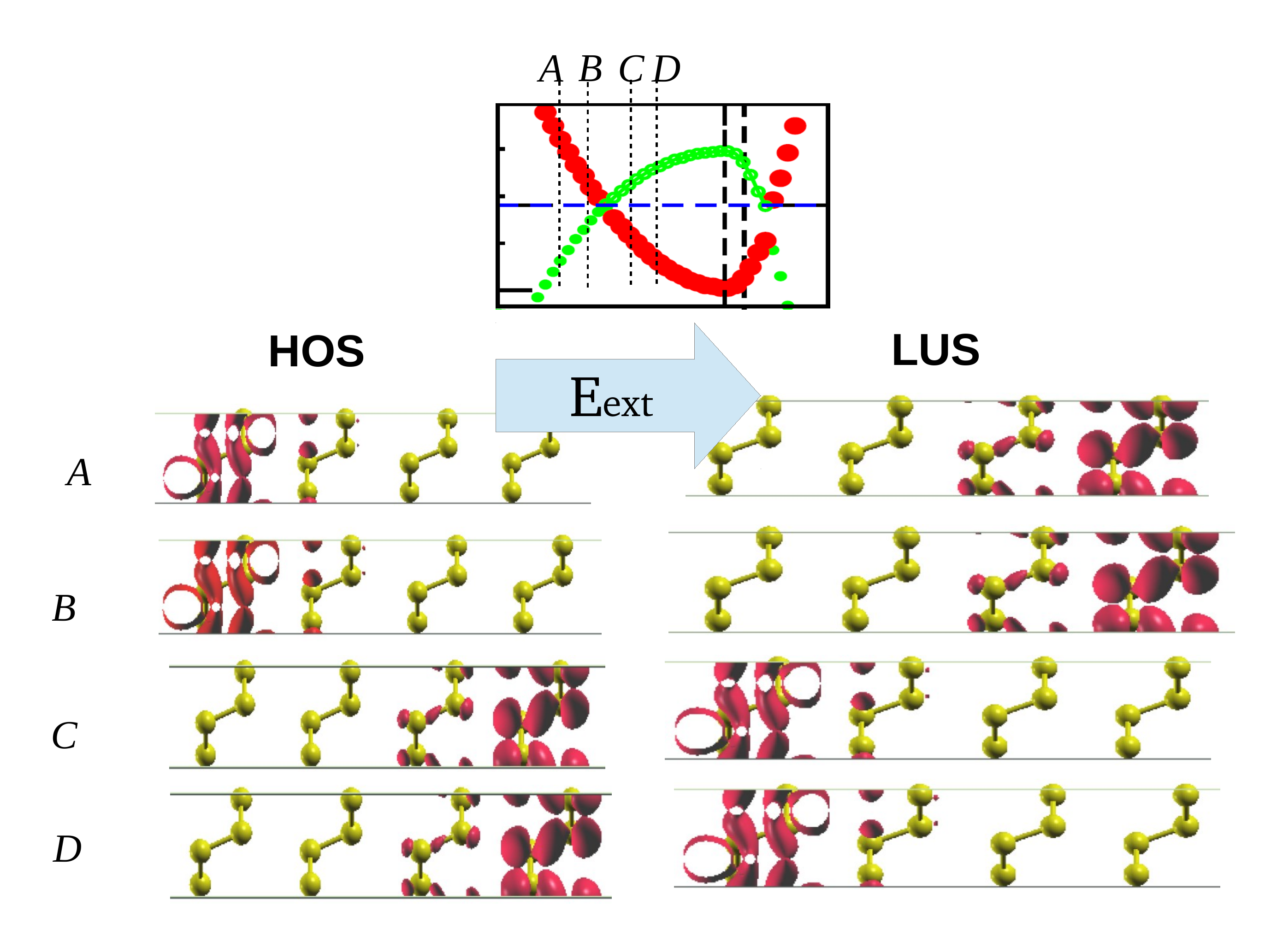}
\caption{{\bf Supporting Figure S5:} The local density of states (LDOS) of highest occupied states (HOS) left) and lowest unoccupied states (LUS) (right) at the different k-points around the Dirac point, $\Lambda$, as indicated in the bandstucture (top). The figure shows that band inversion occurs at the $\Lambda$ point.}
\label{S5}
\end{figure*} 

\begin{figure*}[htbp]
\centering
\includegraphics[width=14.0cm,clip=true]{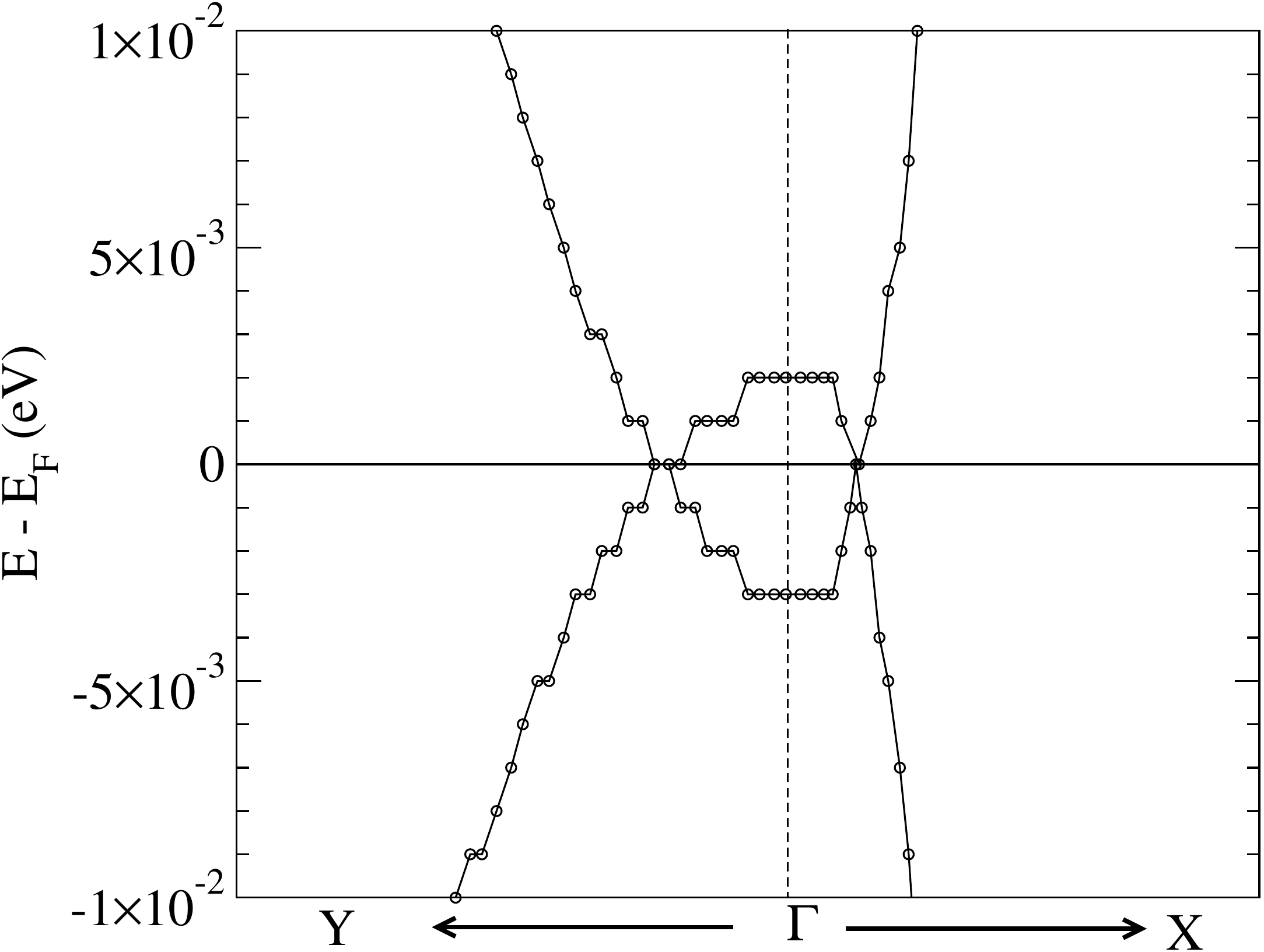}
\caption{{\bf Supporting Figure S6:} The bandstructure of 12L-BP above the critical field, i.e., $E_{\rm ext}$= 0.26 V/\AA.}
\label{S6}
\end{figure*} 

\begin{figure*}[htbp]
\centering
\includegraphics[width=14.0cm,clip=true]{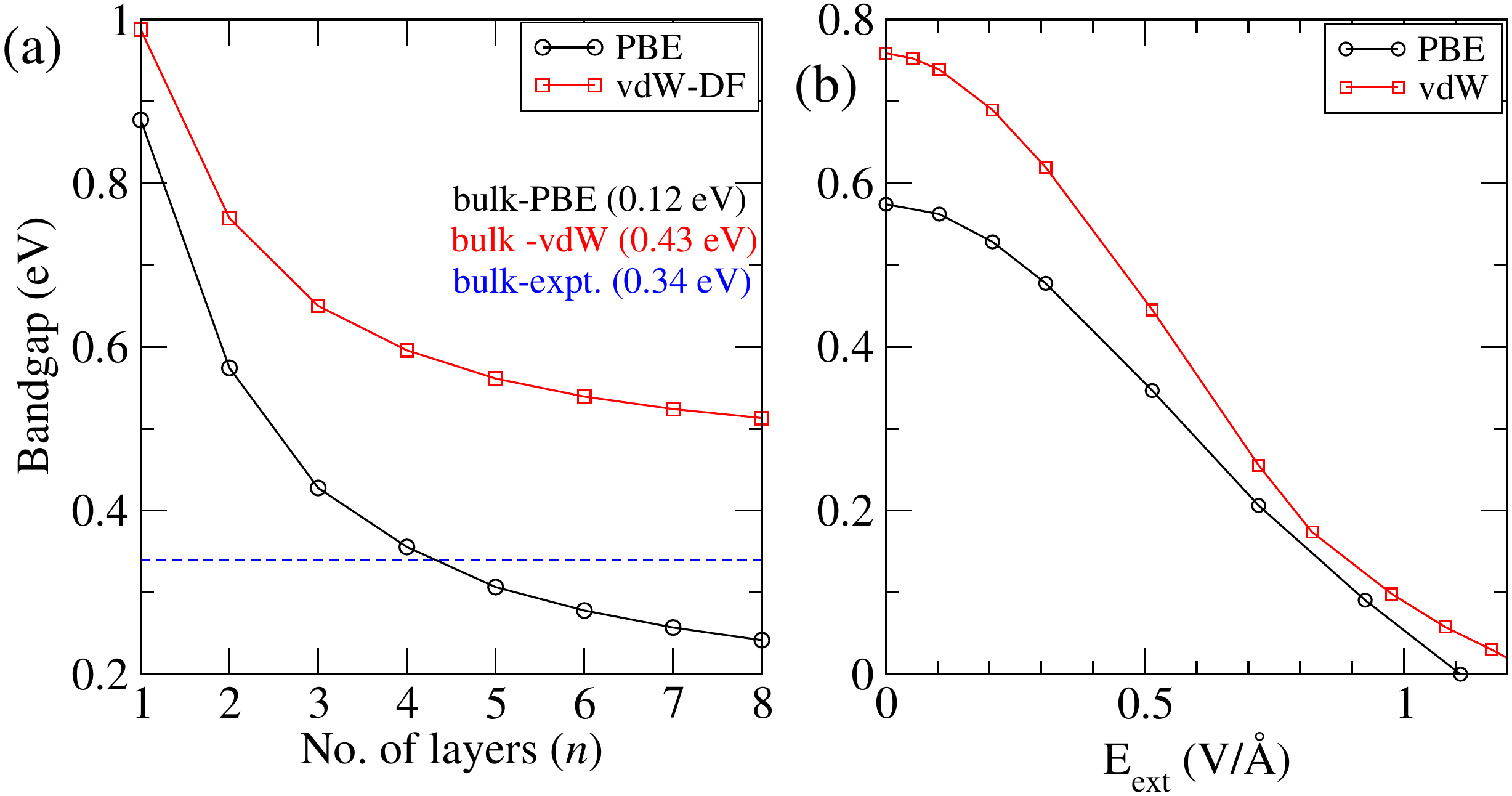}
\caption{{\bf Supporting Figure S7:} Comparison of electronic structure for different functionals. (a) The variation of the band-gaps calculated within PBE(black line) and vdW-corrected(red line) exchange correlation functionals, plotted as a function of the number of layers. Note that the PBE band gap of bulk BP is 0.12 eV, while the vdW-corrected band gap is 0.34 eV. The dashed line denotes the value of the experimental band gap (0.34 eV) of bulk BP. (b) The variation of the band gap of bilayer BP, calculated within PBE(black line) and vdW-corrected(red line) functionals, plotted against the applied field strength.}
\label{S7}
\end{figure*} 

\begin{figure*}[htbp]
\centering
\includegraphics[width=14.0cm,clip=true]{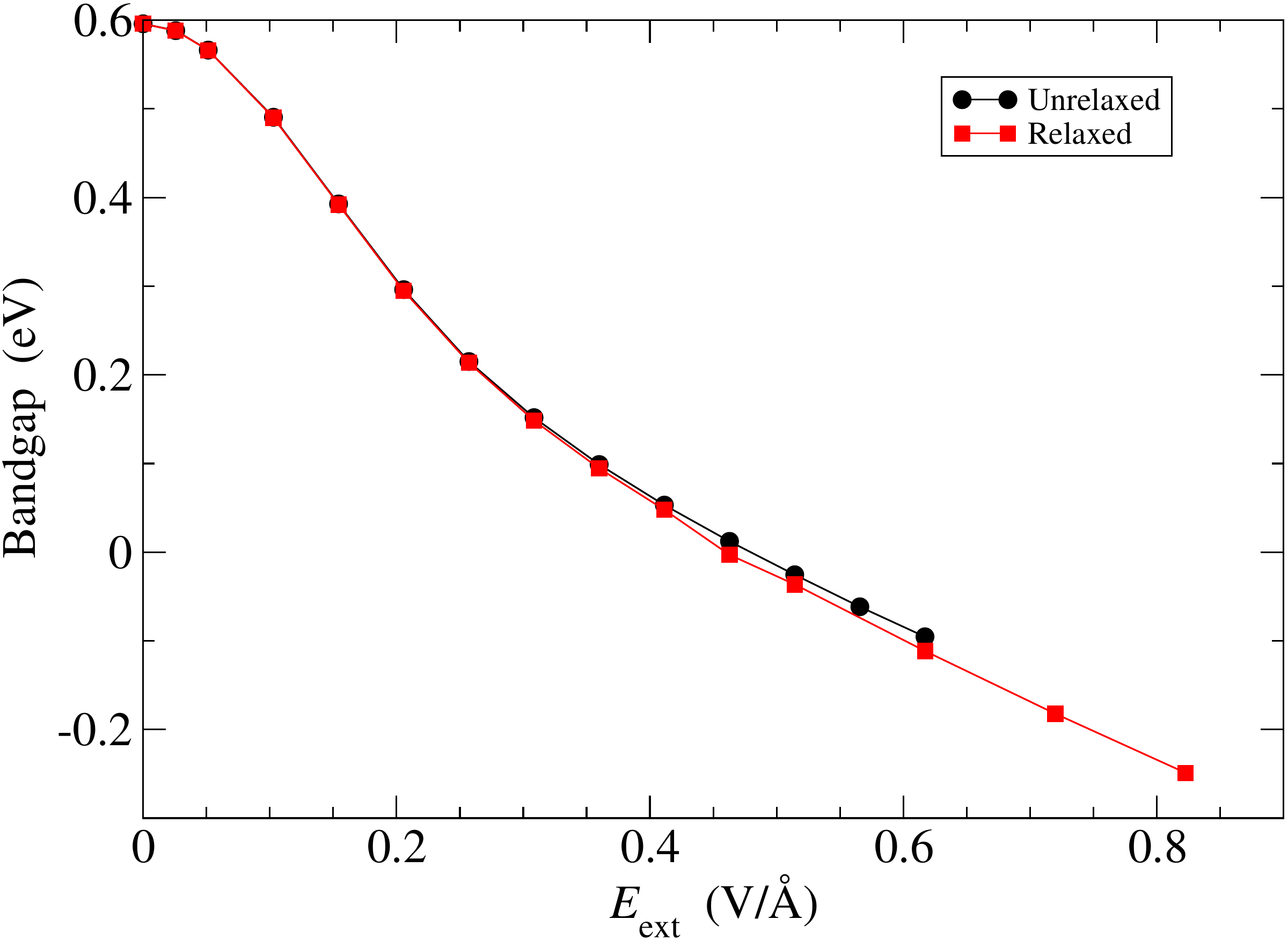}
\caption{{\bf Supporting Figure S8:} The bandgap of 4L-BP as a functional of $E_{\rm ext}$ using the unrelaxed (black) and relaxed (red) coordinates under an applied field $E_{\rm ext}$.}
\label{S8}
\end{figure*} 



\end{document}